\documentclass[reqno]{article}
\usepackage{amsfonts}

\usepackage{epsfig}
\usepackage{amsmath,amssymb}

\title{The Relativistic Particle and its d-brane Cousins}
\bigskip
\author{Vesselin G. Gueorguiev \\
\small \it Department of Physics and Astronomy,\\[-1.mm]
\small \it Louisiana State University, Baton Rouge, LA 70803\\[-1.mm]
\small \it email: vesselin@phys.lsu.edu}
\date{}

\begin{document}

\maketitle
\bigskip

\begin{abstract}
We study properties of classical reparametrization-invariant matter systems,
mainly the relativistic particle and its $d$-brane generalization. The
corresponding matter Lagrangian naturally contains background interaction
fields, such as a 1-form field, analogous to the electromagnetic vector
potential, and a metric tensor. In order to make the theory free of background
fields and prepare for quantum theory of fields, we discuss the field
Lagrangians consistent with the gauge symmetries presented in the equations of
motion for the matter.

\end{abstract}

\section{Introduction}

Probing and understanding physical reality goes through a classical
interface that shapes our thoughts in classical causality chains.
Therefore, understanding of the essential mathematical constructions in
classical mechanics and classical field theory is important, even though
quantum mechanics and quantum field theory are regarded as more
fundamental than their classical counterparts. Two approaches, the
Hamiltonian and the Lagrangian, are very useful in physics \cite {Kilmister
1967, Goldstain 1980,Gracia and Josep,Nikitin-string theory,Carinena
1995}. In general, there is a transformation that relates these two
approaches. For a reparametrization-invariant theory, however, there are
problems in changing to the Hamiltonian approach \cite{Goldstain 1980,
Gracia and Josep, Nikitin-string theory, Rund 1966,Lanczos 1970}.

Fiber bundles provide the mathematical framework for classical mechanics,
field theory, and even quantum mechanics if viewed as a classical field
theory. Parallel transport, covariant differentiation, and gauge symmetry
are very important structures \cite{Pauli 1958} associated with fiber
bundles. When asking what structures are important to physics we should
also ask why one fiber bundle should be more ``physical'' than another, why
the ``physical'' base manifold seems to be a four-dimensional Lorentzian
manifold \cite{Borstnik and Nielsen, van Dam and Ng, Sachoglu 2001}, and
how one should construct an action integral for a given fiber bundle 
\cite{Kilmister 1967, Carinena 1995, Feynman 1965, Gerjuoy and Rau, Rivas
2001}. Starting with the tangent or cotangent bundle seems natural because
these bundles are related to the notion of a classical point-like matter.
Since we accrue and test our knowledge via experiments that involve
classical apparatus, the physically accessible fields should be generated
by matter and should couple with matter as well. Therefore, the matter
Lagrangian should contain the fields, not their derivatives, with which
classical matter interacts \cite{Dirac 1958}.

We study the properties of reparametrization-invariant matter
systems, mainly the relativistic particle \cite{Rund 1966, Pauli 1958,
Feynman 1965, Landau and Lifshitz} and its extended object ($ d $-brane)
generalization. We try to find the answer to the question: ``What is the
Lagrangian for matter?'' The corresponding matter Lagrangian naturally
contains background interaction fields, such as a 1-form field, analogous
to the electromagnetic vector potential, and a metric tensor. We discuss
the guiding principles for construction of field Lagrangians. Due to the
limited space available, we will not discuss the ``non-relativistic''
limit, Klein-Gordon equation, relativistic mass-shell equation, and Dirac
equation here, these topics are covered in ref. \cite{VGG Varna 2002}.

In section two we consider an example of reparametrization-invariant action,
the Lagrangian for a relativistic particle. In the third section we argue in
favor of first order homogeneous Lagrangians. In section four we consider a
possible generalization to the $ D $-dimensional extended objects ($ d $-branes).
Section five contains a review of the field Lagrangians relevant for the
interaction fields. Our conclusions and discussions are in section six.

\section{The Matter Lagrangian for The Relativistic Particle}

From everyday experience we know that localized particles move with a
finite 3D speed. In an extended 4D configuration space-time, when time is
added as a coordinate ($ x^{0}=ct $), particles move with a constant
4-velocity. The 4-velocity is constant because of its definition $ v^{\mu}=
dx^{\mu}/d\tau $ that uses the invariance of the proper time ($ \tau $)
defined via the metric tensor ($ g_{\mu \nu} $) $ d\tau ^{2}= g_{\mu
\nu}dx^{\mu}dx^{v} $. In this case, the action for a massive relativistic
particle has a nice geometrical meaning; it is the distance along the
trajectory \cite{Pauli 1958}: 
\begin{equation}
S_{1}=\int d\tau L_{1}(x,v)=\int d\tau \sqrt{g_{\mu \nu}v^{\mu}v^{\nu}},
\quad \sqrt{g_{\mu \nu}v^{\mu}v^{\nu}}\rightarrow 1\Rightarrow
S_{1}=\int d\tau. \label{S1}
\end{equation}
However, for massless particles, such as photons, the length of the
4-velocity is zero ($ g_{\mu \nu}v^{\mu}v^{\nu}=0 $). Thus one has to use a
different Lagrangian to avoid problems due to division by zero. The
appropriate `good' action is \cite{Pauli 1958}: 
\begin{equation}
S_{2}=\int L_{2}(x,v)d\tau =\int g_{\mu \nu}v^{\mu}v^{\nu}d\tau.
\label{S2}
\end{equation}
Notice that the Euler-Lagrange equations obtained from $ S_{1} $ and $ S_{2} $ are
equivalent, and both are equivalent to the geodesic equation as well: 
\begin{equation}
\frac{d}{d\tau}\vec{v}=D_{\vec{v}}\vec{v}=v^{\beta}\nabla _{\beta}\vec{v}
=0,\quad v^{\beta}\left(\frac{\partial v^{\alpha}}{\partial x^{\beta}}-\Gamma _{\beta \gamma}^{\alpha}v^{\gamma}\right) =0.
\label{geodesic equations}
\end{equation}

In general relativity the Levi-Civita connection $ \nabla _{\beta} $, with
Christoffel symbols $ \Gamma _{\beta \gamma}^{\alpha} $, preserves the length
of the vectors ($ \nabla g(\vec{v},\vec{v})=0 $) \cite{Pauli 1958}.
Therefore, these equivalences are not surprising because the Lagrangians in 
( (\ref{S1}) and (\ref{S2})) are functions of the preserved
length $ g(\vec{v},\vec{v})=\vec{v}^{2} $. However, the parallel transport for
a general connection $ \nabla _{\beta} $ may not preserve the length of the
vectors.

The equivalence between $ S_{1} $ and $ S_{2} $ is very robust. Since $ L_{2} $ is
a homogeneous function of order $ 2 $ with respect to $ \vec{v} $, the
corresponding Hamiltonian function ($ h=v\partial L/\partial v-L $) is exactly
equal to $ L $ ($ h(x,v)=L(x,v) $). Thus $ L_{2} $ is conserved, and so is the
length of $ \vec{v} $. Any homogeneous Lagrangian of order $ n\neq 1 $ is
conserved because $ h=(n-1)L $. When $ dL/d\tau =0 $, then one can show that the
Euler-Lagrange equations for $ L $ and $ L^{\prime}=f\left(L\right) $ are
equivalent under certain restrictions on $ f $. This is an interesting type of
equivalence that applies to homogeneous Lagrangians. It is different from
the usual equivalence $ L\rightarrow L^{\prime}=L+d\Lambda /d\tau $ or the
more general equivalence discussed in ref. \cite{Hojman and Harleston}. Any
solution of the Euler-Lagrange equation for $ L^{\prime}=L^{\alpha} $ would
conserve $ L=L_{1} $ since $ h^{\prime}=(\alpha-1)L^{\alpha} $. All these
solutions are solutions of the Euler-Lagrange equation for $ L $ as well;
thus $ L^{\alpha}\subset L $. In general, conservation of $ L_{1} $ is not
guaranteed since $ L_{1}\rightarrow L_{1}+d\Lambda /d\tau $ is also a
homogeneous Lagrangian of order one equivalent to $ L_{1} $. This suggests
that there may be a choice of $ \Lambda $, a ``gauge fixing'', so
that $ L_{1}+d\Lambda /d\tau $ is conserved even if $ L_{1} $ is not. The above
discussion applies to a general homogeneous Lagrangian.

\section{Homogeneous Lagrangians of First Order}

Suppose we don't know anything about classical physics, which is mainly
concerned with trajectories of point particles in some space $ M $, but we are
told we can derive it from a variational principle if we use the right
action integral $ S=\int Ld\tau $. By following the above example we wonder:
``should the smallest `distance' be the guiding principle?'' when
constructing $ L$. If yes, ``How should it be defined for other field
theories?'' It seems that a reparametrization-invariant theory can provide
us with a metric-like structure \cite{Rund 1966}, and thus a possible link
between field models and geometric models \cite{Rucker 1977}.

In the example of the relativistic particle, the Lagrangian and the trajectory
parameterization have a geometrical meaning. In general, however,
parameterization of a trajectory is quite arbitrary for any observer. Leaving aside
recent hints for quantum space-time from loop quantum gravity and other
theories\footnote{If there is a smallest time interval that sets a space-time
scale, then this would imply a discrete space-time structure since there may not be
any events in the smallest time interval. The Planck scale is often considered to be
such an essential scale \cite{Magueijo and Smolin}.}, we ask: ``Should there be any
preferred trajectory parameterization in a smooth 4D space-time?'' and ``Aren't we
free to choose the standard of distance (time, using natural units $ c=1 $)?'' If
so, then \textit{we really have a smooth continuous manifold and our theory should
not depend on the choice of parameterization}.

If we look at the Euler-Lagrange equations: 
\begin{equation}
\frac{d}{d\tau}\left(\frac{\partial L}{\partial v^{\alpha}}\right) =
\frac{\partial L}{\partial x^{\alpha}}, \label{Euler-Lagrange equations}
\end{equation}
we see that any homogeneous Lagrangian of order $ n $ ($ L(x,\alpha \vec{v})=
\alpha ^{n}L(x,\vec{v}) $) provides a reparametrization invariance ($ \tau
\rightarrow \tau /\alpha, \vec{v}\rightarrow \alpha \vec{v} $). Next, note
that the action $ S $ involves an integration that is a natural structure for
orientable manifolds ($ M $) with an $ n $-form of the volume. Since a
trajectory is a one-dimensional object, then what we are looking at is an
embedding $ \phi:\Bbb{R}^{1}\rightarrow M$. This means that we push forward the
tangential space $ \phi _{*}:T(\Bbb{R}^{1})=\Bbb{R}^{1}\rightarrow T(M)$, and
pull back the cotangent space $ \phi ^{*}:T(\Bbb{R}^{1})=\Bbb{R} ^{1}\leftarrow
T^{*}(M)$. Thus a 1-form $ \omega $ on $ M $ that leaves in $ T^{*}(M) $ ($ \omega
=A_{\mu}\left(x\right) dx^{\mu} $) will be pulled back on $ \Bbb{R}^{1} $ ($ 
\phi ^{*}(\omega) $) and there it is proportional to the volume form
on $ \Bbb{R}^{1} $ ($ \phi ^{*}(\omega)=A_{\mu}\left(x\right)
(dx^{\mu}/d\tau)d\tau \sim d\tau $), allowing us to integrate $ \int \phi
^{*}(\omega) $ : 
\[
\int \phi ^{*}(\omega)=\int Ld\tau =\int A_{\mu}\left(x\right) v^{\mu}d\tau. 
\]

Therefore, by selecting a 1-form $ \omega =A_{\mu}\left(x\right)
dx^{\mu} $ on $ M $ and using $ L=A_{\mu}\left(x\right) v^{\mu} $ we are actually
solving for the embedding $ \phi:\Bbb{R}^{1}\rightarrow M $ using a chart
on $ M $ with coordinates $ x:M\rightarrow \Bbb{R}^{n}$. The Lagrangian obtained
this way is homogeneous of first order in $ v $ with a very simple dynamics. The
corresponding Euler-Lagrange equation is $ F_{\nu \mu}v^{\mu}=0 $ where $ F $ is a
2-form ($ F=dA $); in electrodynamics this is the Faraday's tensor. If we relax
the assumption that $ L $ is a pulled back 1-form and assume that it is just a
homogeneous Lagrangian of order one, then we find a reparametrization-invariant
theory that may have an interesting dynamics.

\subsection{Pros and Cons About Homogeneous Lagrangians of First Order}

Some of the good things about a theory with a first order homogeneous
Lagrangian are:

\begin{itemize}
\item[(1)] Any Lagrangian $ L(x,\frac{dx}{dt}) $ gives rise to a
reparametrization-invariant Lagrangian\footnote{It is an open question whether
there is an equivalence of the corresponding Euler-Lagrange equations.} by
enlarging the space to an extended space-time: $ L(x,\frac{dx}{dt})\rightarrow
L(x,\frac{dx}{dt})\frac{dt}{d\tau} $ \cite {Goldstain 1980}.

\item[(2)] There is a reparametrization invariance of the action $ S=\int
L(x,\frac{dx}{d\tau})d\tau $.

\item[(3)] Parameterization-independent path-integral quantization since the
action $ S $ is reparametrization invariant.

\item[(4)] The reparametrization invariance may help in dealing with
singularities \cite{Kleinert 1989}.

\item[(5)] It is easily generalized to $ D $-dimensional extended objects
($ d $-branes), which is the subject of the next section. Most of the features
listed are more or less self-evident.
\end{itemize}

The list of bad things about a theory with a first order homogeneous
Lagrangian includes:

\begin{itemize}
\item[(1)] There are constraints among the Euler-Lagrange equations
since $ \det \left(\frac{\partial ^{2}L}{\partial v^{\alpha}\partial 
\vee ^{\beta}}\right) =0$, \cite{Goldstain 1980}.

\item[(2)] It follows that the Legendre transformation ($ T\left(M\right)
\leftrightarrow T^{*}\left(M\right) $), which exchanges
coordinates $ (x,v)\leftrightarrow (x,p)$, is problematic \cite{Gracia and
Josep}.

\item[(3)] There is a problem with the canonical quantization approach
since the Hamiltonian function is identically ZERO ($ h\equiv 0 $) \cite
{Nikitin-string theory}.
\end{itemize}

Constraints among the equations of motion are not an insurmountable problem
since there are procedures for quantizing such theories \cite{Nikitin-string
theory, Dirac 1958a, Teitelboim 1982, Henneaux and Teitelboim, Sundermeyer
1982}. For example, instead of using $ h\equiv 0 $ one can use some of the
constraint equations available, or a conserved quantity, as the Hamiltonian
for the quantization procedure \cite{Nikitin-string theory}. Changing
coordinates $ (x,v)\leftrightarrow (x,p) $ seems to be difficult, but it may be
resolved in some special cases by using the assumption that a
gauge $ \Lambda $ has been chosen so that $ L\rightarrow L+ \frac{d\Lambda}
{d\tau}=L^{\prime}=const $. We would not discuss the above-mentioned
quantization troubles since they are outside of the scope of this paper. A
different approach is under investigation, for more details
see ref. \cite{VGG Varna 2002}.

\subsection{Canonical Form of the First Order Homogeneous Lagrangians}

By now, we hope that the reader is puzzled, as we are, about the answer to
the following question: ``What is the general mathematical expression for first order
homogeneous functions?'' Below we define what we mean by the 
\textit{canonical form of the first order homogeneous Lagrangian} and why
we prefer such a mathematical expression.

First, note that any symmetric tensor of rank $ n $ ($ S_{\alpha _{1}\alpha
_{2}...\alpha _{n}}=S_{[\alpha _{1}\alpha _{2}...\alpha _{n}]} $,
where $ [\alpha _{1}\alpha _{2}...\alpha _{n}] $ is an arbitrary permutation of
the indexes) defines a homogeneous function of order $ n $ ($ S_{n}(\vec{v},...,
\vec{v})=S_{\alpha _{1}\alpha _{2}...\alpha _{n}}v^{\alpha _{1}}....v^{\alpha
_{n}} $). The symmetric tensor of rank two is denoted by $ g_{\alpha \beta} $.
Using this notation, the canonical form of the first order homogeneous
Lagrangian is defined as: 
\begin{equation}
L\left(\vec{x},\vec{v}\right) =\sum_{n=1}^{\infty}
\sqrt[n]{S_{n}\left(\vec{v},...,\vec{v}\right)}=A_{\alpha}v^{\alpha}+
\sqrt{g_{\alpha\beta}v^{\alpha}v^{\beta}}+...\sqrt[m]{S_{m}\left(\vec{v},...,
\vec{v}\right)}.
\label{canonical form}
\end{equation}

Whatever is the Lagrangian for matter, it should involve interaction fields
that couple with the velocity $ \vec{v} $ to a scalar. Thus we must
have $ L_{matter}\left(\vec{x},\vec{v};Fields\right) $. When the matter action
is combined with the action for the interaction fields ($\mathcal S= \int
\mathcal{L} dV$), we obtain a full 
\textit{background independent theory}. Then the corresponding Euler-Lagrange
equations contain ``dynamical derivatives'' on the left hand side and sources
on the right hand side: 
\[
\partial _{\gamma}\left(\frac{\delta \mathcal{L}}{\delta (\partial
_{\gamma}\Psi ^{\alpha})}\right) =\frac{\delta \mathcal{L}}{\delta \Psi
^{\alpha}}+\frac{\partial L_{matter}}{\partial \Psi^{\alpha}}. 
\]

The advantage of the canonical form of the first order homogeneous
Lagrangian (\ref{canonical form}) is that each interaction field, which is
associated with a symmetric tensor, has a unique matter source that is a
monomial in the velocities: 
\begin{equation}
\frac{\partial L}{\partial S_{\alpha _{1}\alpha _{2}...\alpha _{n}}}=\frac{1}{n}
\left(S_{n}(\vec{v},...,\vec{v})\right) ^{\frac{1-n}{n}}v^{\alpha
_{1}}....v^{\alpha _{n}}. \label{sources}
\end{equation}

There are many other ways one can write first-order homogeneous functions 
\cite{Rund 1966}. For example, one can consider the following
expression $ L\left(\vec{x},\vec{v}\right) =\left(h_{\alpha
\beta}v^{\alpha}v^{\beta}\right) 
\left(g_{\alpha \beta}v^{\alpha}v^{\beta}\right) ^{-1/2} $ where $ h $ and $ g $ are
seemingly different symmetric tensors. However, each of these fields
($ h $ and $ g $) has the same source type ($ \sim v^{\alpha}v^{\beta} $): 
\[
\frac{\partial L}{\partial h_{\alpha \beta}}=\frac{L\left(\vec{x},\vec{v}
\right)}{h_{\gamma \rho}v^{\gamma}v^{\rho}}v^{\alpha}v^{\beta},\quad 
\frac{\partial L}{\partial g_{\alpha \beta}}=\frac{L\left(\vec{x},\vec{v}
\right)}{g_{\gamma \rho}v^{\gamma}v^{\rho}}v^{\alpha}v^{\beta}. 
\]
Theories with two metrics have been studied before \cite{Dirac 1979,
Bekenstein 1993}. At this stage, however, we cannot find any good reason why
the same source type should produce different fields. Therefore, we prefer the
canonical form (\ref{canonical form}) for our discussion.

\section{$ D $-dimensional Extended Objects}

In the previous sections, we have discussed the classical mechanics of a
point-like particle as a problem concerned with the embedding $ \phi:\Bbb{R}
^{1}\rightarrow M $. The map $ \phi $ provides the trajectory (the word line)
of the particle in the target space $ M $. In this sense, we are dealing with
a $ 0 $-brane that is a one dimensional object. Although time is kept in mind
as an extra dimension, we do not insist on any special structure associate
with a time flow. We think of an extended object as a manifold $ D $ with
dimension, denoted also by $ D, \dim D=D=d+1 $ where $ d=0,1,2,..$.. In this
sense, we have to solve for $ \phi:D\rightarrow M $ such that some action
integral is minimized. From this point of view, we are dealing with
mechanics of a $ d $-brane. In other words, how is this $ D $-dimensional extended
object submerged in $ M$, and what are the relevant interaction fields? By
using the coordinate charts on $ M $ ($ x:M\rightarrow \Bbb{R}^{n} $), we also
can think of this as a field theory over the $ D $-manifold with a local
fiber $ \Bbb{R}^{m} $. Thus the field $ \vec{\phi} $ is such that $ \phi
^{\alpha}=x\circ
\phi:D\rightarrow M\rightarrow \Bbb{R}^{n}$. Following the point particle
discussion, we consider the space of the $ D $-forms over the manifold $ M $,
denoted by $ \Lambda ^{D}\left(M\right) $, that has
dimension $ \binom{m}{D}=\frac{m!}{D!(m-D)!}$. An element $ \Omega $ in $ \Lambda
^{D}\left(M\right) $ has the form $ \Omega =\Omega _{\alpha _{1}...\alpha
_{m}}dx^{\alpha _{1}}\wedge dx^{\alpha _{2}}
\wedge...dx^{\alpha _{m}}$. We use an arbitrary label $ \Gamma $ to index
different $ D $-forms over $ M, \Gamma =1,2,..., \binom{m}{D}$; thus $\Omega
\rightarrow \Omega ^{\Gamma}=\Omega _{\alpha _{1}...\alpha _{m}}^{\Gamma}
dx^{\alpha _{1}}\wedge dx^{\alpha _{2}}\wedge...dx^{\alpha _{m}}$. Next we
introduce ``\textit{generalized velocity vectors}'' with components $ \omega
^{\Gamma} $ : 
\[
\omega ^{\Gamma}=\frac{\Omega ^{\Gamma}}{dz}=\Omega _{\alpha _{1}...\alpha
_{D}}^{\Gamma}\frac{\partial \left(x^{\alpha _{1}}x^{\alpha
_{2}}...x^{\alpha _{D}}\right)}{\partial (z^{1}z^{2}...z^{D})},\quad
dz=dz^{1}\wedge dz^{2}\wedge...\wedge dz^{D}. 
\]
In the above expression, $ \frac{\partial \left(x^{\alpha _{1}}x^{\alpha
_{2}}...x^{\alpha _{D}}\right)}{\partial (z^{1}z^{2}...z^{D})} $ represents
the Jacobian of the transformation from coordinates $ \{x^{\alpha}\} $ over
the manifold $ M $ to coordinates $ \{z^{a}\} $ over the $ d $-brane. The pull
back of a $ D $-form $ \Omega ^{\Gamma} $ must be proportional to the volume
form over the $ d $-brane: 
\[
\phi ^{*}\left(\Omega ^{\Gamma}\right) =\omega ^{\Gamma}dz^{1}\wedge
dz^{2}\wedge...\wedge dz^{D}=\Omega _{\alpha _{1}...\alpha _{D}}^{\Gamma}
\frac{\partial \left(x^{\alpha _{1}}x^{\alpha _{2}}...x^{\alpha
_{D}}\right)}{\partial (z^{1}z^{2}...z^{D})}dz^{1}\wedge dz^{2}\wedge...
\wedge dz^{D}. 
\]
Therefore, it is suitable for integration over the $ D $-manifold. Thus the
action for $ \phi $ is 
\[
S\left[ \phi \right] =\int_{D}L\left(\vec{\phi},\vec{\omega}\right)
dz=\int_{D}\phi ^{*}\left(\Omega \right) =\int_{D}A_{\Gamma}
(\vec{\phi})\omega ^{\Gamma}dz. 
\]
This is a homogeneous function in $ \omega $ and has reparametrization
(diffeomorphism) invariance with respect to the $ D $-manifold. If we relax
the linearity $ L(\vec{\phi},\vec{\omega}) =\phi ^{*}\left(\Omega 
\right) =A_{\Gamma}(\vec{\phi})\omega ^{\Gamma} $ in $ \vec{\omega} $, then the
canonical expression for the homogeneous Lagrangian is: 
\begin{equation}
L\left(\vec{\phi},\vec{\omega}\right) =\sum_{n=1}^{\infty}\sqrt[n]{S_{n}
\left(\vec{\omega},...,\vec{\omega}\right)}=A_{\Gamma}\omega
^{\Gamma}+\sqrt{g_{\Gamma _{1}\Gamma _{2}}\omega ^{\Gamma _{1}}\omega
^{\Gamma _{2}}}+...\sqrt[m]{S_{m}\left(\vec{\omega},...,\vec{\omega}\right)}. 
\label{canonical d-brane L}
\end{equation}

At this point, there is strong analogy between a point particle and a $ d $-brane.
However, there is a difference in the number of components; $ \vec{x}, \vec{v}$,
and $ \vec{\phi}=\vec{x}\circ \phi $ have the same number of components, but the
``generalized velocity'' $ \vec{\omega} $ has $ \binom{\dim M}{\dim D} $ components
which are Jacobians \cite{Fairlie and Ueno}.

Some familiar Lagrangians include:

\begin{itemize}
\item The Lagrangian for a 0-brane (relativistic point particle in an
electromagnetic field, $\dim D=1 $ and $ \omega ^{\Gamma}\rightarrow
v^{\alpha}=\frac{dx^{\alpha}}{d\tau} $) is: 
\[
L\left(\vec{\phi},\vec{\omega}\right) =A_{\Gamma}\omega ^{\Gamma}+
\sqrt{g_{\Gamma _{1}\Gamma _{2}}\omega ^{\Gamma _{1}}\omega ^{\Gamma _{2}}}
\rightarrow L\left(\vec{x},\vec{v}\right) =qA_{\alpha}v^{\alpha}+
m\sqrt{g_{\alpha\beta}v^{\alpha}v^{\beta}}. 
\]

\item The Lagrangian for a 1-brane (strings, $ \dim D=2 $) \cite{Nikitin-string theory}
is: 
\[
L\left(x^{\alpha},\partial _{i}x^{\beta}\right) =\sqrt{Y^{\alpha 
\beta}Y_{\alpha \beta}}, 
\]
using the notation: 
\[
\omega ^{\Gamma}\rightarrow Y^{\alpha \beta}=\frac{\partial (x^{\alpha},
x^{\beta})}{\partial (\tau,\sigma)}=\det \left(\begin{array}{cc}
\partial _{\tau}x^{\alpha} & \partial _{\sigma}x^{\alpha} \\ 
\partial _{\tau}x^{\beta} & \partial _{\sigma}x^{\beta}
\end{array}
\right) =\partial _{\tau}x^{\alpha}\partial _{\sigma}x^{\beta}-\partial
_{\sigma}x^{\alpha}\partial _{\tau}x^{\beta}. 
\]

\item The Lagrangian for a $ d $-brane has the Dirac-Nambu-Goto term (DNG)
\cite{Pavsic 2001}: 
\[
L\left(x^{\alpha},\partial _{D}x^{\beta}\right) =\sqrt{Y^{\Gamma}Y_{\Gamma}}. 
\]
\end{itemize}

Notice that most of the Lagrangians above, except for the relativistic
particle, are restricted only to gravity-like interactions. In the case of
the charged relativistic particle, the electromagnetic interaction is very
important. Thus similar interaction should be introduced in the string
theory and in the DNG model.

\section{The Background Fields and Their Lagrangians}

The uniqueness of the interaction fields and source types has been essential for
the selection of the matter Lagrangian (\ref{canonical d-brane L}). The first two
terms in the Lagrangian are easily identified as electromagnetic and gravitational
interaction. The other terms are somewhat new. It is not yet clear if these new
terms are real or not, so we will not engage them actively in the following
discussion. At this stage, we have a theory with background fields since we don't
know the equations for the interaction fields. To complete the theory, we
need to introduce actions for these interaction fields.

One way to write the action integrals for the interaction fields $ S_{n} $ in 
(\ref{canonical d-brane L}) follows the case of the $ d $-brane discussion. There,
we have been solving for $ \phi:D\rightarrow M $ by selecting a Lagrangian that is
more than a pull back of a $ d $-form over the manifold $ M$. In a similar way, we
may view $ S_{n} $ as an $ M $-brane field theory, where $ S_{n}:M\rightarrow
S_{n}M $ and $ S_{n}M $ is the fiber of symmetric tensors of rank $ n $ over $ M$.
This approach, however, cannot terminate itself since new interaction fields would
be generated as in the case of $ \phi:D\rightarrow M$. 

Another way assumes that $ A_{\Gamma} $ is a 1-form. Thus we may use the
external algebra structure $\Lambda \left(T^{*}M\right) $ over $ M $ to construct
objects proportional to the volume form over $ M $. For any $ n $-form $(A)$ objects
proportional to the volume form $ \Omega _{Vol} $ can be constructed by using the
external derivative $ d$, multiplication $ \wedge$, and Hodge dual $ * $ operations
in $ \Lambda\left(T^{*}M\right)$. For example, $ A\wedge *A $ and $ dA\wedge *dA $
are forms proportional to the volume form. 

The next important ingredient comes from the symmetry in the matter equation. That is,
if there is a transformation $ A\rightarrow A^{\prime} $ that leaves the matter
equations unchanged, then there is no way to distinguish $ A $ and $ A^{\prime} $.
Thus the action for the field $ A $ should obey the same symmetry (gauge symmetry).

For example, the matter equation for 4D electromagnetic interaction is $ d
\vec{v}/d\tau =F\cdot \vec{v} $ where $ F $ is the 2-form obtained by
differentiation of the 1-form $(A)$ ($ F=dA $), and the gauge symmetry
for $ A $ is $ A\rightarrow A^{\prime}=A+df$. The reasonable terms for a 1-form
field in the field Lagrangian $ \mathcal{L}(A) $ are: $ A\wedge *A, dA\wedge
*dA$, and $ dA\wedge dA$. The first term does not conform with the gauge
symmetry $ A\rightarrow A^{\prime}=A+df $ and the second term $ (dA\wedge dA) $ is a
boundary term since $ dA\wedge dA=d\left(A\wedge dA\right) $ that
gives $ \int_{M}d\left(A\wedge dA\right) =A\wedge dA $ at the boundary of $ M$. This
term is interesting in the quantum Hall effect. Therefore, we are left with a
unique action for electromagnetism: 
\[
S\left[ A\right] =\int_{M}dA\wedge *dA=\int_{M}F\wedge *F. 
\]

For our next example, we look at the terms in the matter equation that
involve gravity. There are two possible choices of matter equation. The first
one is the geodesic equation $ d\vec{v}/d\tau =\vec{v}\cdot \Gamma
\cdot \vec{v} $ where $ \Gamma $ is considered as a connection 1-form that
transforms in the usual way $ \Gamma \rightarrow \Gamma +\partial g $ under
coordinate transformations ($ g $). This type of transformation, however, is
not a `good' symmetry since restricting $ \Gamma \rightarrow \Gamma +\Sigma $ to
transformations $ \Sigma =\partial g, $ such that $ \vec{v}\cdot \Sigma
\cdot \vec{v}=0 $, would mean to select a subset of coordinate systems,
inertial systems, for which the action $ S $ is well defined and
satisfies $ S\left[ \Gamma \right] = $ $ S\left[ \Gamma +\Sigma \right] $.
Selecting a class of observables for the description of the system is not
desirable, so we shall not follow this road.

In general, the Euler-Lagrange equations assume a background observer who
defines a coordinate system. For electromagnetism, this is tolerable
since neutral particles are such privileged observers. In gravity, however,
there is no such observer, and the equation for matter should be
relational. Such an equation is the equation of the geodesic
deviation: $ d^{2}\vec{\xi}/d\tau ^{2}=R\cdot \vec{\xi} $, where $ R $ is a
Lie $ \left(TM\right) $ valued curvature 2-form $ R=d\Gamma +[\Gamma,\Gamma ]$. A
general curvature 2-form is denoted by $ F\rightarrow $ $ \left(F_{\alpha
\beta}\right) _{j}^{i} $. Here, $ \alpha $ and $ \beta $ are related to the
tangential space of the base manifold $ M $. The $ i $ and $ j $ are related to the
fiber structure of the bundle where is given the connection that
defines $ \left(F_{\alpha \beta}\right) _{j}^{i} $. Clearly, the Ricci
tensor $ R $ is a very special curvature because all of its indices are
of $ TM $ type. For that reason, it is possible to contract the fiber degree of
freedom with the base manifold degree of freedom (indices). Thus an action
linear in $ R $ is possible. In general, one needs to consider a quadratic action,
i.e. trace of $ F\wedge *F $ ($ F_{\alpha \beta j}^{i}\wedge *F_{\alpha \beta
i}^{j}$).

Using the symmetries of the Ricci tensor $ R $ ($ R_{\alpha \beta,\gamma \rho}=-R_{\beta \alpha,\gamma \rho}=-R_{\alpha \beta,\rho \gamma}=R_{\gamma
\rho,\alpha \beta} $) we have two possible expressions that can be
proportional to the volume form $ \Omega$. The first expression is present in
all dimensions and is denoted by $ R^{*} $, which means that a Hodge
dual operation has been applied to the second pair of indices ($ R_{\alpha
\beta,*(\gamma \rho)} $). The $ R^{*} $ action seems to be related to the
Cartan-Einstein action for gravity $ S\left[ R\right] =\int R_{\alpha
\beta}\wedge *(dx^{\alpha}\wedge dx^{\beta}) $ \cite{Adak et al 2001}.

The other expression is only possible in a four-dimensional space-time and
involves full anti-symmetrization of $ R $ ($ R_{\alpha [\beta,\gamma ]\rho)} $)
denoted by $ R^{\wedge} $. We have not been able to identify the role of
the $ R^{\wedge} $ yet. Such a term in the action could produce a stabilizing
effect or restore the renormalizability of the four dimensional gravity. If
this happens, then it could be one of the reasons why the spacetime seems to be
four dimensional. However, a statistical argument \cite{Sachoglu 2001} based
on geometric and differential structure of various brane and target spaces
seems to be a better explanation for why we are living in a 4D
space-time. For example, if we make a statistical path-integral-like
estimate using the following generating function: 
\[
Z[S]=\sum_{\dim M=1}^{\infty}\left (\sum_{M-topologies}\left (\sum_{\dim D=1}^{\dim
M}\left (\sum_{D-topologies}\left (\sum_{other\ structures} e^{-S_{m}\left[
\phi:D\rightarrow M;S_{n}\right]-S_{f}\left[ S_{n}\right]} \right) \right)
\right) \right), 
\]
then we may find that the expectation value for the averaged space dimension is 4
because of the infinitely many homeomorphic but not diffeomorphic 4D spaces. In the
expression above $ S_{m}\left[ \phi:D\rightarrow M;S_{n}\right] $ is the action
for matter with $ S_{n} $-type fields, and $ S_{f}\left[ S_{n}\right] $ is the
corresponding action for the interaction fields.

\section{Conclusions and Discussions}

In summary, we have discussed the structure of the matter Lagrangian ($ L $) for
extended objects. Imposing reparametrization invariance of the action $ S $ naturally
leads to a first order homogeneous Lagrangian. In its canonical form, $ L $ contains
electromagnetic and gravitational interactions, as well as interactions that are
not clearly identified yet. If one extrapolates from the strengths of the two known
interactions, then one may suggest that the next terms should be important, if
present at all, at big cosmological scales, such as galactic cluster dynamics. The
choice of the canonical Lagrangian is based on the assumption of one-to-one
correspondence between interaction fields and the type of their sources. If one can
show that any homogeneous function can be written in the canonical form suggested,
then this would be a significant step in our understanding of the fundamental
interactions. Note that an equivalent expression can be considered as well: $
L=A_{\alpha}(\vec{x},\vec{v})v^{\alpha} $. This expression is simpler, and is
concerned with the structure of the homogeneous functions of order zero $
A_{\alpha}(\vec{x},\vec{v}) $. If one is going to study the new interaction fields
$S_n, n>2$, then the guiding principles for writing field Lagrangians, as discussed
in the examples of electromagnetism and gravity, may be useful. It would be
interesting to apply the outlined constructions to general relativity by considering
it as a $ 3 $-brane in a $ 10 $ dimensional target space ($ g_{\alpha
\beta}:M\rightarrow S_{2}M $).

{\bf Acknowledgments.} The author acknowledges helpful discussions with
Professors R. Hymaker, L. Smolinsky, A. R. P. Rau, R. F. O'Connell, P. Kirk,
J. Pullin, C. Torre, J. Baez, P. Al. Nikolov, E. M. Prodanov, G. Dunne, and L.
I. Gould. The author is also thankful to Professor J. P. Draayer at the LSU
Department of Physics and Astronomy, Dr. Joe Abraham at the LSU Writing
Center, and the organizers of the of the First Advanced Research Workshop on
Gravity, Astrophysics and Strings at The Black Sea in Kiten, Bulgaria, who
provided the opportunity for this research and its publication. The author
acknowledges the financial support from the Department of Physics and
Astronomy and the Graduate School at the Louisiana State University, the U. S.
National Science Foundation support under Grant No. PHY-9970769 and
Cooperative Agreement No. EPS-9720652 that includes matching from the
Louisiana Board of Regents Support Fund.

\end{document}